\documentclass{article}
\usepackage{amsmath,epsfig}
\title{Comparison of Zn$_{1-x}$Mn$_x$Te/ZnTe multiple-quantum wells 
and quantum dots by 
below-bandgap photomodulated reflectivity}

\author{P. J. Klar and D. Wolverson\\
School of Physics, University of East Anglia\\
Norwich NR4 7TJ, UK\\
\and
D. E. Ashenford and B. Lunn\\
Department of Engineering Design and Manufacture\\
 University of Hull,
Hull HU6 7RX, UK\\
\and
Torsten Henning\\
Department of Physics\\ 
University of G\"oteborg and Chalmers University of Technology\\
S-41296 G\"oteborg, Sweden}
\date{17 June 1996\\
\texttt{Semicond. Sci. Technol. \textbf{11}, 1863-1872 (1996)\\
cond-mat/9710008}}

\begin{document}
\maketitle
\begin{abstract}

Large-area high density patterns of quantum dots with a 
diameter of 200\,nm have been 
prepared from a series of four 
Zn$_{0.93}$Mn$_{0.07}$Te/ZnTe multiple quantum well structures 
of different well width (40\,\AA{}, 60\,\AA{}, 80\,\AA{} and 100\,\AA{}) 
by electron beam lithography followed by 
Ar$^+$-ion beam etching. Below-bandgap photomodulated reflectivity 
spectra of the 
quantum dot samples and the parent heterostructures were then 
recorded at 10\,K and the 
spectra were fitted to extract the linewidths and the energy 
positions of the excitonic 
transitions in each sample. The fitted results are compared to
calculations of the transition 
energies in which the different strain states in the samples 
are taken into account. We 
show that the main effect of the nanofabrication process is a 
change in the strain state of 
the quantum dot samples compared to the parent heterostructures. 
The quantum dot 
pillars turn out to be freestanding, whereas the heterostructures 
are in a good 
approximation strained to the ZnTe lattice constant. The lateral 
size of the dots is such 
that extra confinement effects are not expected or observed.
\end{abstract}

\section{Introduction}

A subgroup of the dilute magnetic semiconductors (DMS) is formed by
alloys 
consisting 
of a II-VI host material with Mn$^{2+}$ ions substituted for some of the

group II cations (e.g. 
Zn$_{1-x}$Mn$_x$Te, Cd$_{1-x}$Mn$_x$Te and Zn$_{1-x}$Mn$_x$Se). 
The most prominent feature of these Mn-containing DMS materials 
is the s,p-d exchange interaction between the $S = 5/2$ 
d-electrons of the Mn$^{2+}$ ions and the free charge carriers in the 
conduction and valence 
bands, which leads to magnetic splittings at low temperatures of the 
order of 100\,meV 
(for a review of these properties see [1, 2]). In recent years, 
heterostructures such as 
multiple quantum wells (MQWs) containing DMS materials have been 
grown very 
sucessfully by several groups. Progress has been made in the 
understanding of the 
properties of Mn at the heterointerfaces [3, 4]. It is now 
possible to use this knowledge to 
make detailed studies, by magnetic field experiments, of 
aspects of epitaxial growth such 
as asymmetric interfaces [5] and donor distributions in MQW samples [6],

demonstrating 
that DMS systems are ideal model systems for the study of semiconductor 
heterostructures. These II-VI DMS alloys are also of interest since 
they are closely related 
to the II-VI wide-bandgap alloys such as Zn$_{1-x}$Cd$_x$Se used in 
blue-green optoelectronic 
devices.

The free carriers and the excitons in heterostructures are confined 
in one dimension 
forming a two-dimensional (2D) quantum system. It is of interest 
from the viewpoint of 
fundamental physics, as well as that of applications in device 
manufacturing, to reduce the 
dimensionality of quantum systems further, to one-dimensional (1D) 
systems such as 
quantum wires or even zero-dimensional (0D) systems such as 
quantum dots (Q-dots). 
There are several approaches by which one can achieve this, 
including the direct growth of 
1D or 2D systems or the controlled etching of heterostructures. 
Some examples of direct 
growth techniques of III-V and II-VI 1D and 0D structures are 
the growth of 
microcrystallites [7, 8], growth on patterned substrates [9, 10], 
cleaved edge overgrowth 
[11, 12], strain modulation by using sputtered stressors [13, 14] 
and self-organized growth 
[15, 16]. The controlled etching techniques are all based on a 
lithography process to 
create an etch pattern followed by an etching process. Several 
techniques for etching II-VI 
materials have been used: ion beam etching (IBE) [17], wet 
etching [18, 19], reactive ion 
etching (RIE) [20, 21] or combinations of the three techniques. 
Similar techniques have 
been used for III-V materials (for an overview. see [22]). 
The smallest Q-dot sizes 
achieved by etching of II-VI materials are of the order of 30\,nm 
in diameter, which is 
close to the size at which additional confinement effects are 
significant [23]. Nevertheless, 
the study of Q-dots of bigger diameter can yield valuable 
information about changes in the 
strain state of the specimen or about the damage caused by the 
nanofabrication process. 
Two main techniques are used to study the strain contributions 
in nanostructures. These 
are X-ray diffractometry [24-26] and photomodulated reflectivity 
[27-30] and both reveal 
that, in most cases, a relaxation towards a freestanding structure 
takes place in quantum 
dots and quantum wires. The damage induced by the nanofabrication 
process has been 
investigated by time-resolved and continuous wave photoluminescence 
spectroscopy 
[31-33] and Raman spectroscopy of phonons [34]. So far, only a few 
1D or 0D DMS 
structures have been prepared, all in the Cd$_{1-x}$Mn$_x$Te/CdTe 
system [28, 35-37] .  

We report here on a study of the strain states in a series of four 
200\,nm diameter 
Zn$_{1-x}$Mn$_x$Te/ZnTe MQWs quantum dot samples (and their parent 
heterostructures) 
using below-bandgap photomodulated reflectivity (BPR). In the 
BPR technique, the 
reflectivity of the sample is modulated by  laser light whose 
energy is below the band gap 
of the semiconductor material; BPR was first reported as a new 
technique by 
Bhimnathwala 
and Borrego [38], although it was already applied before by 
Rockwell et al. [39]. 
We have established that BPR spectra contain the same spectroscopic 
information as 
conventional photomodulated reflectivity spectra [40]. 
We have tried to gain some insight 
into the modulation mechanism of BPR and have shown that 
it is likely to result from the 
excitation of deep levels in the semiconductor material [41].

\section{Experimental details}

The four MQW samples were grown by molecular beam epitaxy (MBE) 
on (001)-oriented 
GaSb substrates. Each heterostructure consists of a 1000\,\AA{} 
buffer layer followed by 10 
ZnTe quantum wells embedded between Zn$_{0.93}$Mn$_{0.07}$Te 
barriers. The four samples are 
of different quantum well width 40\,\AA{}, 60\,\AA{}, 80\,\AA{} and
100\,\AA{}
respectively and the barrier 
thickness is 150\,\AA{} in all cases. Further details of the MBE 
growth can be found in [40].

The nanofabrication of the quantum dots was carried out in eight steps. 
First, the specimen 
was cleaned with acetone in an ultrasonic bath and rinsed with
isopropanol. 
In the second 
step, it was spin-coated with a two-layer resist consisting of a 140\,nm

bottom layer of 
copolymer P(MMA-MAA) and a 50\,nm top layer of 950k PMMA. Afterwards, the 
electron beam lithography was carried out with a JEOL JBX 5DII system 
with CeB$_6$ cathode 
using an acceleration voltage of 50\,kV, a final aperture of 60\,$\mu$m,

a working distance of 
14\,mm and a beam current of 1\,nA. The patterned area covered $3.2
\times 3.2$\,mm$^2$ on the 
specimen. We exposed $200 \times 200$\,nm$^2$ squares with a nominally 
homogeneous dose of 
220\,$\mu$C/cm$^2$. These squares were arranged on a square lattice 
with a lattice constant of 
400\,nm. The diameter of the beam and the proximity effect result in a 
circular exposure 
profile under these conditions. In the fourth step, the exposed sample 
was developed in a 
solution of 20\,mL isopropanol and 2\,mL distilled water for 60\,s. 
Afterwards, a 30\,nm Ti 
film was deposited on the resist mask, and after subsequent lift-off 
in warm acetone, a 
pattern of circular Ti dots with a diameter of about 200\,nm to 220\,nm 
was obtained. This 
pattern was transferred onto the semiconductor using Ar$^+$ IBE under 
normal incidence 
with a relatively low accelaration voltage of 200\,V and a current density of 
0.16\,mA/cm$^2$. With these parameters, about 20\,min were requried to 
etch through the 
heterostructure to a depth of about 30\,nm. In the final, eighth step
the 
Ti etch mask was 
removed with hydrofluoric acid. After each process step, the samples
were 
inspected by 
scanning electron microscopy (SEM). 

\begin{figure}[t]
\begin{center}
\epsfig{file=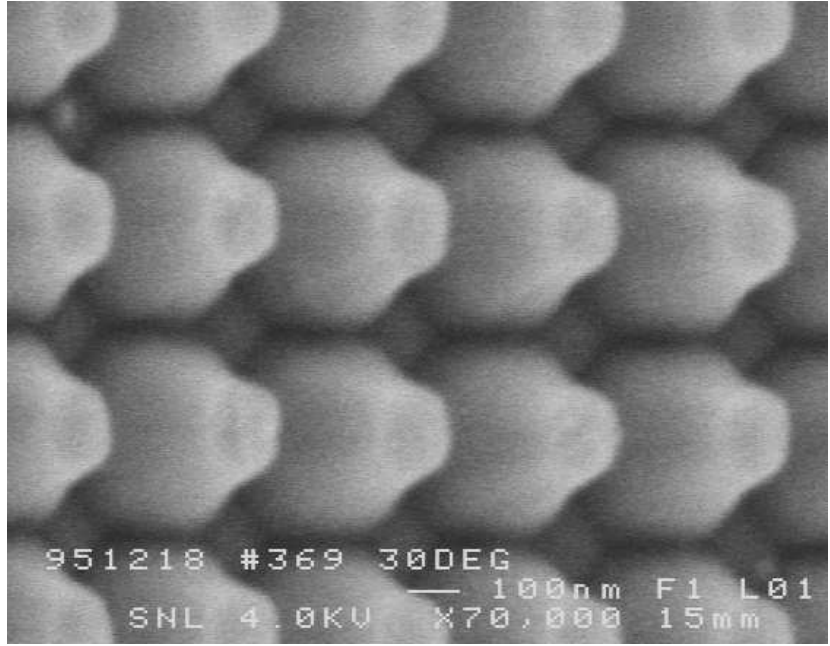,width=0.9\textwidth}
\end{center}
\caption{\label{fig1}Scanning electron microscope image of
the quantum dots 
prepared from the 
Zn$_{0.93}$Mn$_{0.07}$Te/ZnTe 
multiple quantum well sample with 80\,\AA{} well width. The sample 
is tilted by 30$^\circ$ and the acceleration voltage is 4\,kV. }
\end{figure}

Figure 1 shows an SEM image of the high density pattern of 200\,nm
Q-dots 
prepared on 
the Zn$_{0.93}$Mn$_{0.07}$Te/ZnTe heterostructure with 80\,nm quantum 
well width. The ten 
quantum wells are situated in the top two-thirds of the dots. 
The image was taken with the 
specimen tilted by 30$^\circ$ and the Ti mask has been removed. 
The inspection with the 
SEM showed that the size of the dots as well as the period of the 
dot lattice are both 
homogeneous throughout the whole $3.2 \times 3.2$\,mm$^2$ area. 
We have obtained a high density 
of dots, close to the attempted ratio of the area covered by dots 
to the etched area of $1:3$. 
There are only a few regions with defects in the dot pattern 
(not shown in figure 1). 
These originate mainly from macroscopic defects such as scratches 
and dust particles 
already on the sample before the nanofabrication process and were 
not caused by the 
process itself. The shape of the individual dots shows an 
inclination of the side walls, 
which is typical for IBE etched samples [42]. There are two 
main reasons for this 
inclination in IBE: erosion of the metal mask and redeposition 
of etched materials. A 
stronger erosion of the edges of the Ti mask compared to the 
central mask region occurs 
due to the angle dependence of the Ti etching rate. 
This sucessive reduction of the metal 
mask is then transferred into the semiconductor as the 
etching continues resulting in 
different etching depths. Because of the low acceleration 
voltage of the Ar$^+$ ions, the 
sputtered atoms are of low energy and are simply redeposited 
on opposite surfaces. The 
inclination observed in our samples is about 25$^\circ$ 
which is
larger than the 
10$^\circ$ to 15$^\circ$ 
observed by other groups in II-VI materials [17, 18], using 
comparable acceleration 
voltages in the IBE process. This might be caused by a 
sensitivity of the IBE process to 
the symmetry of the dot pattern as observed for other 
etching processes like RIE [43]. In 
our relatively dense pattern, the redeposition of the 
sputtered atoms might be easier and, 
therefore, the shielding of the Ti mask more effective 
with increasing etch depth. It is 
known from other studies [31, 33] that the IBE process 
introduces a damage layer on the 
sidewalls of the dots of approximately 30\,nm width. 
It was further shown that the damage 
layer can be removed successfully by anodic oxidation. 

For the optical measurements, the specimen was cooled 
down to 10\,K within an  Oxford 
Instruments closed cycle He cryostat (CCC1204). 
The BPR experiments were performed 
in near-backscattering geometry using the white light of 
a tungsten lamp as the probe 
light. The light passing through a pinhole placed in front 
of the tungsten lamp was 
focussed to a 2\,mm diameter image of the pinhole on the 
sample by using a concave 
mirror in a 2$f$ set-up. The reflected light was collected 
with a lens and focussed by a 
second lens onto the slit of a 0.5\,m Spex single spectrometer 
and detected with an S20 
photomultiplier. A shutter in front of the pinhole allowed the 
probe light to be switched on 
and off. Chopped modulation light was provided by a 10\,mW HeNe 
laser of wavelength 
633\,nm (1.96\,eV) which is below the excitonic bandgap of the 
ZnTe (2.380\,eV) at low 
temperatures. The modulation frequency was typically 280\,Hz. 
The modulation light was 
focussed onto the same spot on the sample as the white probe light. 

The output of the photomultiplier was connected to a two-channel 
Stanford SR530 Lock-in 
amplifier and a Keithley 177 Microvolt DMM Voltmeter. 
The AC and DC components 
of the output of the photomultiplier were measured by the 
lock-in amplifier and the 
voltmeter respectively. The first channel of the lock-in 
amplifier was used to detect the 
component of the photomultiplier signal in phase with the modulation 
laser and the second 
channel was used to read the 90$^\circ$ out-of-phase component. 
Both lock-in channels and the 
voltmeter signal were readable by computer via an IEEE interface. 
The following 
measurement cycle allowed a measurement of the relative change in
reflectivity 
$dR/R$ at 
each spectrometer position. When the shutter was open (so that the 
modulation laser light 
and the probe light were on the sample) the voltmeter recorded the 
reflectivity signal $R$ 
plus a DC background signal $B$ whilst the two lock-in channels 
recorded the AC changes 
of the reflectivity plus background signals, 
$dR_\mathrm{in} + dB_\mathrm{in}$  and $dR_\mathrm{out} +
dB_\mathrm{out}$, 
respectively. 
When the shutter was closed (so that only the modulation laser light 
was on the sample) 
the three background signals $B$, $dB_\mathrm{in}$ and $dB_\mathrm{out}$
respectively 
were recorded. The values 
for the relative changes of the reflectivity 
$dR_\mathrm{in}/R$ and $dR_\mathrm{out}/R$ as well as the reflectivity 
signal $R$ were then computed. The time needed to complete one
measurement 
cycle was 
about 10\,s using a time-constant of 1\,s for the lock-in and several 
averages of the 
instrument readings. The spectra were recorded in the range of
5000\,\AA{} to 5300\,\AA{}
in 1\,\AA{}
steps, which was the resolution of the spectrometer (corresponding to an

energy 
resolution of about 0.5\,meV).

\section{Fit and model calculations}

All modulated reflectivity $dR/R$ lineshapes are based on the
following expression
\begin{equation}
\frac{dR}{R}=\alpha\,d\varepsilon_1 + \beta\,d\varepsilon_2,
\end{equation}
where $\alpha$ and $\beta$ are the Seraphin coefficients 
and $d\varepsilon_1$ and $d\varepsilon_2$ are the changes in the real 
and imaginary parts of the complex dielectric function $\varepsilon$
[44]. 
The complex dielectric 
function in the vicinity of an excitonic or electronic transition is 
given by the sum of a 
background contribution, $\varepsilon_\mathrm{b}$, 
(which is assumed to be independent of the perturbation) 
and the contribution of the transition itself, 
$\varepsilon_\mathrm{t}$, which changes in response to an external 
perturbation (in our case, the electric field of the modulation
laser):
\begin{align}
\varepsilon &=  \varepsilon_\mathrm{b} + \varepsilon_\mathrm{t} \\
d\varepsilon &=  d\varepsilon_1 = i\,d\varepsilon_2 \approx
      d\varepsilon_\mathrm{t1} + i\,d\varepsilon_\mathrm{t2} = 
      d\varepsilon_\mathrm{t} \notag.
\end{align}
All lineshapes used in the fitting of $dR/R$ spectra are based on
functional forms 
$\varepsilon_\mathrm{t}(E_\mathrm{g},I,\Gamma)$ 
where $E_\mathrm{g}$, $I$ and $\Gamma$ are the energy position, the intensity
and the 
linewidth of the transition. 
So far, no functional forms appropriate for PR lineshapes have been 
proposed for 
transitions in 0D or 1D systems. However, the quantum dots which we 
have prepared are 
still of sizes where additional confinement effects are negligible [23],

which justifies the 
use of the same lineshapes in the fits of the spectra for the quantum
dot 
and the MQW 
samples. The choice of the lineshapes is described in great detail 
elsewhere [40] and is 
briefly summarized in the following. The change 
$d\varepsilon_\mathrm{t}$  for allowed excitonic transitions in 
the quantum well can be described by the first derivative of a complex 
Lorentzian 
lineshape with respect to transition energy $E_\mathrm{g}$
\begin{equation}
d\varepsilon_\mathrm{t} \propto I \,\frac{E-E_\mathrm{g}-i\Gamma}%
{E-E_\mathrm{g}+i\Gamma},
\end{equation}
where $E$ is the photon energy. To fit the signals of bulk-like barrier 
and buffer transitions 
we have used a third-derivative-like change of the dielectric function
\begin{equation}
d\varepsilon_\mathrm{t} \propto \frac{I}{\left(%
E-E_\mathrm{g}-i\Gamma\right)^{5/2}},
\end{equation}
which was proposed to hold for a single critical point of a 
direct bandgap with a three-dimensional 
parabolic density of states [45]. The $dR/R$ lineshape of each
oscillator 
is then 
given by equation (1) and has the four fit parameters 
$\alpha$, $\beta$, $\Gamma$ and $E_\mathrm{g}$. The $dR/R$ spectrum 
is fitted by a sum of single-oscillator lineshapes (one for each
transition) using a 
least-square procedure.

We have carried out calculations of the optical transition energies
similar 
to the ones 
described earlier [40]. The confining potentials for light holes, 
heavy holes and electrons 
were calculated taking strain effects into account. The strain shifts 
in the conduction band 
and the heavy-hole and light-hole valence bands were determined for 
each layer using the 
expressions derived by Bir and Pikus [46]:
\begin{align}
\Delta{}E_\mathrm{el}&=2a_\mathrm{c}\,\frac{C_{11}-C_{12}}{C_{11}}\,%
\varepsilon_\mathrm{par},\\
\Delta{}E_\mathrm{lh}&=2a_\mathrm{v}\left(\frac{C_{11}-C_{12}}{C_{11}} -
   b\,\frac{C_{11}+2
  C_{12}}{C_{11}}\right)\,\varepsilon_\mathrm{par},\notag\\
\Delta{}E_\mathrm{hh}&=2a_\mathrm{v}\left(\frac{C_{11}-C_{12}}{C_{11}} +
  b\,\frac{C_{11}+2
  C_{12}}{C_{11}}\right)\,\varepsilon_\mathrm{par}\notag,
\end{align}
where 
$\varepsilon_\mathrm{par}=\frac{a_\mathrm{par}-a_0}{a_0}$
is the in-plane strain in each layer, 
$a_\mathrm{par}$ and $a_0$ are the actual and 
the equilibrium lattice constants respectively of the layer at the
measurement temperature
and $C_{11}$ and $C_{12}$ are the elastic stiffness constants. 
The energies of the single particle states in the resulting two valence 
band potentials and 
the conduction band potential were then determined by a one-dimensional 
transfer matrix 
technique. The contribution of the exciton binding energy was determined

separately. 
Here, we have calculated the exciton binding energies using a model
based 
on a variational 
approach which takes the single particle wavefunctions of the electron 
and the hole into 
account [47, 48] rather than the analytical model used in our earlier
work (the
model of Mathieu et al.
[49]). The major reason for the change to the variational method was 
that the model used 
previously is only applicable to the lowest excitonic quantum well 
transitions whereas the 
new model can be applied also to excited states if the wavefunctions 
of the single-particle 
states are known. The exciton binding energies for the e1hh1 and the 
e1lh1 transition 
derived with the two models do differ: using the variational approach, 
we obtain binding 
energies of the heavy-hole exciton of about 20\,meV and of the light 
hole exciton of about 
17\,meV compared to values of about 15\,meV for both exciton binding 
energies using the 
simple model. It will be seen below that this uncertainty in the exciton

binding energy does 
not affect the interpretation of the experimental results.

The parameters in the calculation are those given earlier [40] with some

exceptions. We 
use a chemical valence band offset (VBO) of 30\%, which was determined 
via magnetic 
field experiments [50] on other pieces of the same set of MQW samples 
and by assuming 
the ratio of the absolute deformation potentials
$a_\mathrm{c}$:$(-a_\mathrm{v})$ 
to be 90:10, as calculated 
previously [51]. This combination of parameters describes the net 
band alignment in our 
MQW samples accurately, but it should be pointed out that there are 
other combinations 
of VBO and $a_\mathrm{c}$:$a_\mathrm{v}$ ratio which lead to the same
net band alignment: 
in other words, the 
VBO cannot be determined uniquely until $a_\mathrm{c}$:$a_\mathrm{v}$ 
is measured. To account for the bowing 
of the excitonic bandgap of Zn$_{1-x}$Mn$_x$Te as a function of 
Mn content, $x$, at low 
temperature we use $E_\mathrm{g} = 2.3805$\,eV for the unstrained
excitonic bandgap of 
ZnTe at 2\,K 
[52] and the following linear approximation for the excitonic 
bandgap dependence [53]:
$E_\mathrm{g}(x) = 2.376\,\textrm{eV} + 0.820\,\textrm{eV} \,x
\quad\textrm{for}\quad  
0 < x < 0.6$. There are several other 
reported measurements of the excitonic bandgap of Zn$_{1-x}$Mn$_x$Te 
as a function of 
composition [54-56] which all differ from each other and the dependence
given in [53]. 
We choose the result given above since it is consistent with our 
measured energy of 
the barrier heavy-hole transition in the MQW samples with x = 7\% 
on the assumption that 
the MQWs are strained to the ZnTe lattice constant.

For the following discussion it is important to consider two strain 
situations: that in which 
the heterostructure is strained to the ZnTe lattice constant and that 
where the 
heterostructure is strained to the Zn$_{0.93}$Mn$_{0.07}$Te barrier
lattice constant. 
\begin{figure}
\begin{center}
\epsfig{file=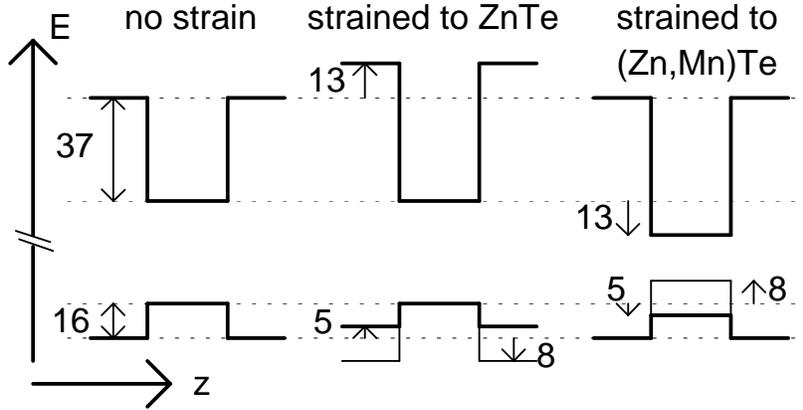,width=0.9\textwidth,%
bbllx=86,bblly=456,bburx=508,bbury=670,clip=}
\end{center}
\caption{\label{fig2}Possible band alignments for a quantum
well 
in the heterostructure. Left:  
isolated layers; centre: sample strained to the ZnTe lattice constant; 
right: sample strained 
to the Zn$_{0.93}$Mn$_{0.07}$Te barrier lattice constant. 
The numbers indicate the potential depth 
for the isolated layers (assuming a VBO of 30\%) and the strain shifts 
for the two strain 
situations assuming a ratio of $a_\mathrm{c}$:$(-a_\mathrm{v})$ of 90:10
(see text). 
The arrows indicate the 
direction of the strain shifts. }
\end{figure}
Figure 2 shows 
the resulting band alignments for a single quantum well for these 
two strain situations 
(centre and right) and the situation for isolated layers 
(for example without strain effects; left). The 
numbers indicate the strain shifts in meV (centre and right) and the 
band offsets in meV 
(left) using the parameters given above. Several aspects are noteworthy
here. 
Firstly, the 
strain shifts modify the potential profiles for the two strain 
situations compared with the 
isolated layer case: the electron potential and the light-hole 
potential become deeper by 
13\,meV and 8\,meV respectively and the heavy-hole potential 
becomes shallower by 
5\,meV. However, the resulting potential profiles for the two 
strain situations are the same 
and the potentials are only arranged differently on the absolute 
energy scale, as depicted in 
figure 2. This is the case because, to a first approximation, the 
deformation potentials and 
elastic constants have been assumed to be the same for ZnTe and 
Zn$_{0.93}$Mn$_{0.07}$Te and 
the in-plane strain $\varepsilon_\mathrm{par}$ only changes sign, 
that is, it is negative (compressive strain) in the 
Zn$_{0.93}$Mn$_{0.07}$Te barriers when the sample is strained to the
ZnTe 
lattice constant and it is 
positive (tensile strain) in the ZnTe wells when the sample is 
strained to the 
Zn$_{0.93}$Mn$_{0.07}$Te barriers. Secondly, in terms of excitonic 
transitions, it turns out that the 
light-hole transitions are more sensitive to the strain effects 
than the heavy hole transitions, 
because the strain shifts in the strained layer for the light-hole 
band and the electron band 
are in opposite directions on the abolute energy scale 
whereas the heavy-hole band shifts 
in the same direction as the electron band. Because the 
potential profiles are the same in 
the two strain situations, we can assume that the exciton 
binding energies of the quantum 
well transitions are also the same. Using this 
assumption, the differences between the 
excitonic transitions in the two strain situations 
are fully determined by the relative shifts 
of the potentials on the absolute energy scale, 
giving shifts of 21\,meV for light-hole 
transitions and 8\,meV for the heavy-hole transitions. Furthermore, we
expect the heavy-hole transition (e1hh1) 
to be the lowest transition when the heterostructure is strained to 
the ZnTe lattice constant and the light-hole transition (e1lh1) 
to be the lowest transition 
when the heterostructure is strained to the Zn$_{0.93}$Mn$_{0.07}$Te
lattice constant.

\section{Results}

The BPR spectra of the Zn$_{0.93}$Mn$_{0.07}$Te/ZnTe MQW with quantum 
well width 100\,\AA{} 
and the 200\,nm diameter Q-dot sample based on it are shown in
figure~3.
\begin{figure}
\begin{center}
\epsfig{file=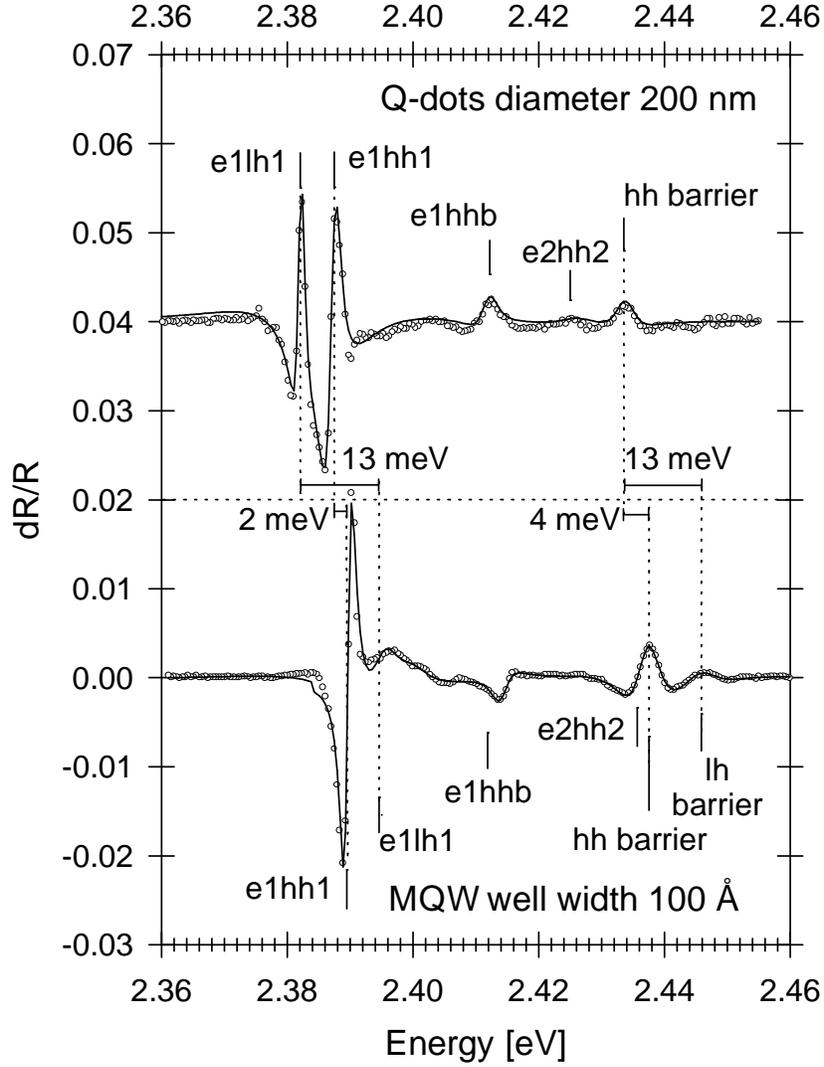,width=0.9\textwidth}
\end{center}
\caption{\label{fig3}Comparison of the below-bandgap photomodulated 
reflectivity spectra of the 
200\,nm diameter quantum dot sample and the parent 
Zn$_{0.93}$Mn$_{0.07}$Te/ZnTe multiple 
quantum well sample with 100\,\AA{} well width. The spectra were 
taken at 10\,K using a 
HeNe laser (1.96\,eV) as the modulation source. The arrows mark the 
fitted energy 
positions of the main excitonic transitions. The dotted lines and the 
solid horizontal bars 
indicate the energy shifts of the e1hh1 and e1lh1 quantum well excitons 
and of the barrier 
transitions as obtained from the fit results. }
\end{figure}
The Q-dot 
spectrum is shifted vertically by 0.04 units for clarity. We use only 
the in-phase 
components of $dR/R$ to represent the spectra (in all figures) because
the 
90$^\circ$ out-of-phase 
components are featureless. This is to be expected since, firstly, the 
sample is thin 
compared to the wavelength of the probe light and, secondly, the 
modulation period is 
very long compared with the time-scales of the relaxation processes 
in the samples. Both 
BPR spectra in figure~3 have  the same signal to noise ratio and 
order of magnitude, 
which makes it easy to compare them. This is not the case for the 
unmodulated reflectivity 
spectra, which for the Q-dots are a factor of ten weaker (and show a
different 
non-excitonic background) than for the MQWs due to the diffraction of 
the probe light by the 
Q-dot pattern. Therefore, BPR is a better tool than reflectivity for 
the study of the 
nanofabricated samples. In the figure the open circles are the measured 
values and the 
full curves are the fits to the spectra. The arrows above and below 
the spectra mark the 
fitted energy positions of the transitions in the Q-dot spectrum 
and the MQW spectrum 
respectively; the fitted positions of the ZnTe buffer signal are 
not indicated. The spectrum 
of the MQW sample was fitted with seven oscillators. We used 
third-derivative lineshapes 
(equation~(4)) for the ZnTe buffer transition and the light-hole 
and heavy-hole barrier 
transitions denoted as hh barrier and lh barrier respectively. 
First derivatives with respect 
to energy position of a complex Lorentzian lineshape (equation~(3)) 
were used for the four 
quantum well related transitions. Three of these can be assigned easily:

these are the 
lowest quantum well heavy-hole (e1hh1) and light-hole (e1lh1) 
excitons and the second 
allowed heavy-hole excitonic transition (e2hh2). In assigning the 
fourth quantum well 
related transition,  the magnetic field splitting in the Faraday
geometry of the 
$\sigma^+$ and the $\sigma^-$ 
components of the signal provides useful evidence. This splitting is 
comparable to the 
splitting of the heavy-hole barrier exciton, indicating that the 
signal corresponds to a 
heavy-hole related transition which may either be a forbidden 
transition involving the first 
electron state and a high-index heavy-hole state of the quantum well 
(probably e1hh3) or 
an indirect transition between a quantum well electron state and the 
heavy-hole barrier 
state. Our preliminary conclusion is that the signal corresponds to the 
spatially-indirect 
well-to-barrier transition (e1hhb) since a signal of this type has been 
observed in a 
Cd$_{0.97}$Mn$_{0.03}$Te/CdTe MQW sample of similar structure and
similar 
confinement 
situation by low-temperature photoluminescence excitation experiments
[57]. 
The 
spectrum of the Q-dot sample was fitted with six oscillators, again
using 
first-derivative 
lineshapes for the four quantum well related transitions, but now only 
two third-derivative 
lineshapes representing only one barrier signal (hh barrier) apart from 
the ZnTe buffer 
signal. We do not observe any additional features in the BPR of the 
Q-dot sample that 
could be related to additional lateral confinement effects in the dots, 
confirming that the 
dot diameter is still in the regime where the Coulomb effects dominate 
over lateral 
confinement effects. It is possible to observe lateral confinement 
effects by 
photomodulated reflectivity experiments in smaller Q-dots, such as 
modulation-doped 
GaAs/Ga$_{1-x}$Al$_x$As Q-dots of 60\,nm diameter at 77\,K [58] and 
Cd$_{1-x}$Mn$_x$Te/CdTe 
Q-dots of 30-40\,nm even at 300\,K [35]. 

The differences between our spectra can be related to different 
strain states of the Q-dots 
and the MQW. The dotted lines in figure~3 correspond to the energies 
of the lowest 
quantum well transitions (e1hh1 and e1lh1) and to the barrier
transitions 
(e1hh barrier and 
e1lh barrier). The solid horizontal bars on the horizontal dotted line 
indicate the shifts of 
these transitions before and after the nanofabrication process. We find 
that these shifts are 
qualitatively in agreement with our simple considerations of the strain 
shifts for light-hole 
and heavy-hole excitonic transitions between the two strain situations 
of the sample 
strained to the ZnTe buffer for the MQW and the sample strained to the
Zn$_{0.93}$Mn$_{0.07}$Te 
barrier for the Q-dot sample. The above discussion of the strain state 
of the samples 
predicts a strain shift of the light-hole transitions of 21\,meV. In the

analysis of the spectra, 
we find a shift of 13\,meV for the light-hole transitions, whilst for
the 
heavy-hole 
transitions we predict a shift of 8\,meV and observe shifts of 2\,meV
and 
4\,meV. Similar 
behaviour is observed in the other specimens and can be explained fully 
on the assumption 
of two well-defined strain states for the two series of samples, as we 
now show.

\begin{figure}
\begin{center}
\epsfig{file=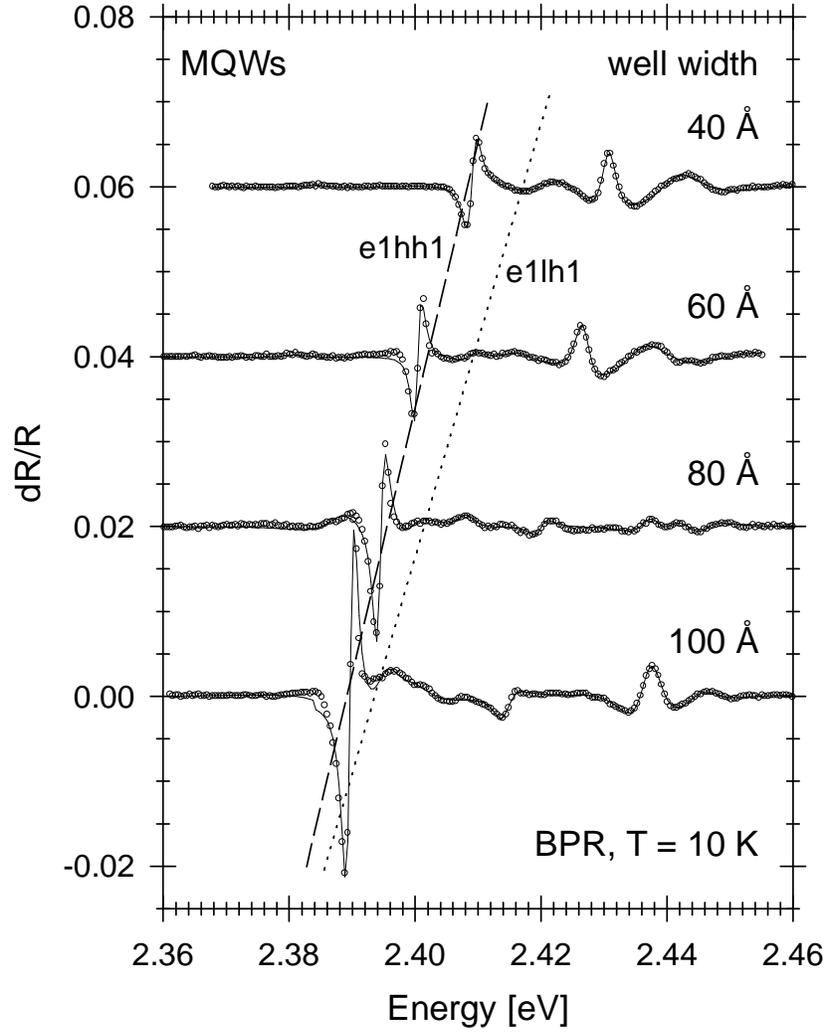,width=0.9\textwidth}
\end{center}
\caption{\label{fig4}Below-bandgap photomodulated
reflectivity 
spectra of the series of four 
Zn$_{0.93}$Mn$_{0.07}$Te/ZnTe multiple quantum well samples of 
different well width. The 
spectra were taken at 10\,K with a HeNe laser as the modulation source. 
The quantum well 
width for each sample is indicated on the right of the figure. 
The dotted and the broken
lines indicate the trend in the splitting between the lowest-energy 
quantum well transitions 
e1lh1 and e1hh1 respectively as a function of well width. }
\end{figure}

\begin{figure}
\begin{center}
\epsfig{file=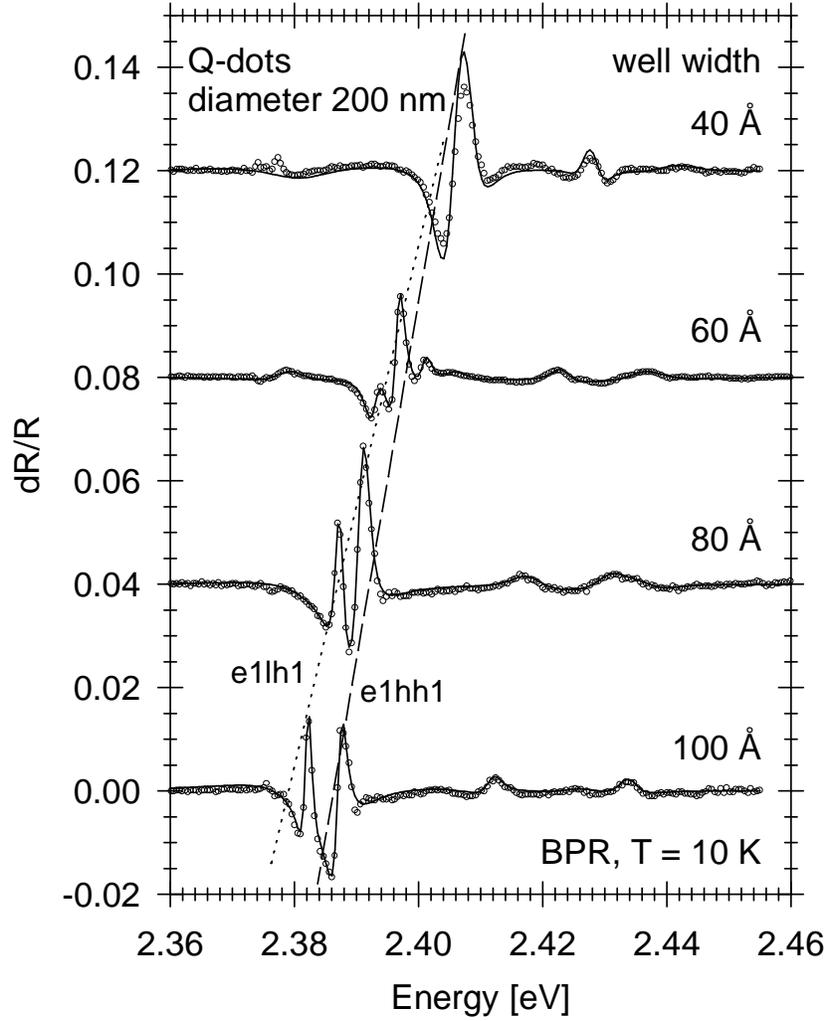,width=0.9\textwidth}
\end{center}
\caption{\label{fig5}Below-bandgap photomodulated reflectivity 
spectra of the series of four 
quantum dot samples with dot diameters of 200\,nm prepared from 
Zn$_{0.93}$Mn$_{0.07}$Te/ZnTe multiple quantum well samples of 
different quantum well width. 
The spectra were taken at 10\,K with a HeNe laser as the modulation
source. 
The well 
width of the parent multiple quantum well is indicated on the right 
of the figure. The 
dotted and broken lines indicate the trend in the splitting between 
the lowest-energy 
quantum well transitions e1lh1 and the e1hh1 respectively as a 
function of well width. }
\end{figure}

Figures~4 and~5 display the measured and fitted spectra of the series of MQW 
samples and of the series of Q-dot samples. The spectra are shifted 
vertically for clarity 
and are arranged in order of increasing quantum well width from top 
to bottom. Again, 
the open circles are the measured values and the full curves are the 
fits. In each figure, a 
dotted line and a broken line indicate the change of the energy 
position of the e1lh1 
transition and the e1hh1 transition respectively. The fits of the 
spectra of the MQW and 
the Q-dots with quantum well width of 80\,\AA{} were the same as 
described for the 
corresponding samples of 100\,\AA{} quantum well width, that is, 
seven oscillators for the 
MQW spectrum and six for the Q-dot spectrum. In the fits of the 
four spectra of the 40\,\AA{} 
and the 60\,\AA{} MQW and Q-dot samples we have not included an 
oscillator for a e2hh2 
transition, thus leaving just six and five oscillators 
respectively to describe the 
experimental spectra. In figure~5, the spectrum of the Q-dot 
sample with 60\,\AA{} quantum 
well width shows two additional features which are situated at 
the energy positions of the 
e1hh1 and the e1lh1 transition of the corresponding MQW sample
(these features are difficult to discern on the scale of figure~5). 
These suggest that we 
have not etched through the whole heterostructure and that the 
lowest quantum well is 
still intact. To account for this in the fitting, we have 
included two more oscillators, fixed 
at the energy positions already determined for the e1lh1 
and e1hh1 transitions of the 
MQW sample. We now focus on the behaviour of the lowest energy 
excitonic transitions 
in the quantum wells, i.e., e1hh1 and e1lh1 and relate this 
behaviour to the strain state of 
the sample. For the MQW samples we expect from the potential 
situation shown in the 
centre of figure~2 that the e1hh1 transition is lower than the 
e1lh1 transition and that, with 
increasing well width, the splitting between the two transitions 
decreases because the 
excitons become less sensitive to the strain-splitting of the
barrier (the changes in the 
quantum confinement with well width are less significant than 
the strain effects in this 
system). This behaviour can be clearly seen in the spectra in 
figure~4 as indicated by the 
convergence of the broken and dotted line. 

From the potential situation for the Q-dots depicted on the 
right of figure~2, we conclude 
that, in the Q-dots, the e1lh1 and the e1hh1 transition will 
behave in the opposite way to 
that described for the MQWs. This agrees well with the behaviour 
displayed in figure~5: 
the e1lh1 transition is lowest in energy in the wells of the Q-dots 
and the splitting between 
the e1hh1 and e1lh1 transitions increases as the quantum well width 
increases, that is, as 
the lowest quantum well transitions become increasingly like those 
in a strained ZnTe 
layer. Another interesting aspect is that in the spectra of the Q-dots, 
we observe an 
enhancement of the intensity of the e1lh1 transition compared to the 
e1hh1 transition, 
which is quite significant for the Q-dot samples of well width
80\,\AA{} and 
100\,\AA{} where the 
intensity ratio ranges from 1:3 to 1:1. A similar effect was observed in 
GaAs/Ga$_{0.7}$Al$_{0.3}$As quantum dots with diameters of 500\,nm, 
400\,nm and 230\,nm and 
was explained by matrix element effects [27] as follows. 
The intensity of a transition is, in 
a first approximation, proportional to the square of the electric
dipole matrix element:
\begin{equation}
I \propto \left| \left\langle c \left| p \right| v \right\rangle \right| ^2,
\end{equation}
where $\left| c \right\rangle$ and
$\left| v \right\rangle$ are the conduction and valence band 
states respectively and $p$ is the 
electric dipole operator. Making use of the degeneracy in 
$m_z$ in the absence of a magnetic 
field, the following expressions for the 
conduction band state and the heavy-hole and light-hole valence 
band states in the 
$\left|J,m_z\right\rangle$ representation hold [59]:
\begin{align}
\left| c\right\rangle &=
\left| \frac{1}{2},\pm\frac{1}{2}\right\rangle = i\,\left| s \right\rangle \\
\left| \nu_\mathrm{hh}\right\rangle &= \left|
    \frac{3}{2},\pm\frac{3}{2}\right\rangle
    =\frac{1}{\sqrt{2}}\left| X+iY\right\rangle\notag\\
\left| \nu_\mathrm{lh}\right\rangle &= \left|
    \frac{3}{2},\pm\frac{1}{2}\right\rangle
    =\frac{1}{\sqrt{6}}\left| X-iY\right\rangle-\sqrt{\frac{2}{3}}
    \left| Z \right\rangle\notag.
\end{align}
For an epitaxially grown heterostructure, the probe light is 
normally incident parallel to the 
growth direction $z$, thus the dipole operator only has components in
the $xy$ plane, leading 
to an intensity ratio of 1:3 between the lowest-energy light-hole 
and heavy-hole 
transitions. In the case of quantum dots, the probe light can 
penetrate into the sample via 
the sidewalls, thus having a component of the dipole moment 
in the $z$-direction which, 
from equation~(7), increases the intensity of the light-hole
transition. 
(Note that in figure~7 
the intensity ratio appears to decrease with the well width 
and the two transitions 
approach each other, indicating that matrix element effects can only 
explain qualitatively 
the behaviour of the intensity ratio; however, the present experimental 
evidence is 
insufficient to draw firm conclusions.)

To verify the proposed model for the strain states of the samples 
quantitatively, we have 
carried out calculations of the energies of the observed transitions 
in each case. 
Comparisons of the transition energies obtained from the fitting 
of the spectra with the 
results of the calculations are shown in figures~6 and~7. 
\begin{figure}
\begin{center}
\epsfig{file=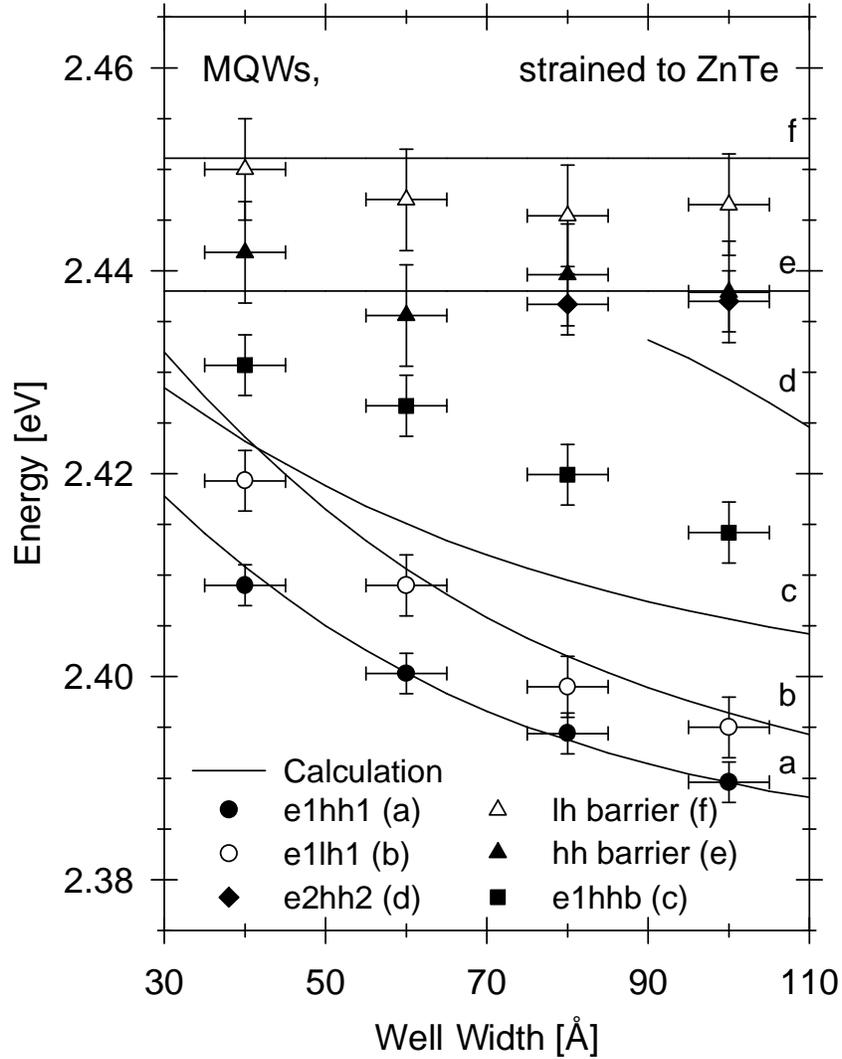,width=0.9\textwidth}
\end{center}
\caption{\label{fig6}Comparison of the calculations (full curves
labelled a to f) 
for the case in which 
the heterostructure is strained to the ZnTe lattice constant 
with the experimental results 
(symbols) obtained for the excitonic transition energies in the four 
Zn$_{0.93}$Mn$_{0.07}$Te/ZnTe multiple quantum well samples. 
The horizontal error bars result 
from the assumption of an uncertainty of 5\,\AA{} (2 monolayers) 
in the quantum well width 
and the vertical bars represent the fitted linewidth of each transition. }
\end{figure}
\begin{figure}
\begin{center}
\epsfig{file=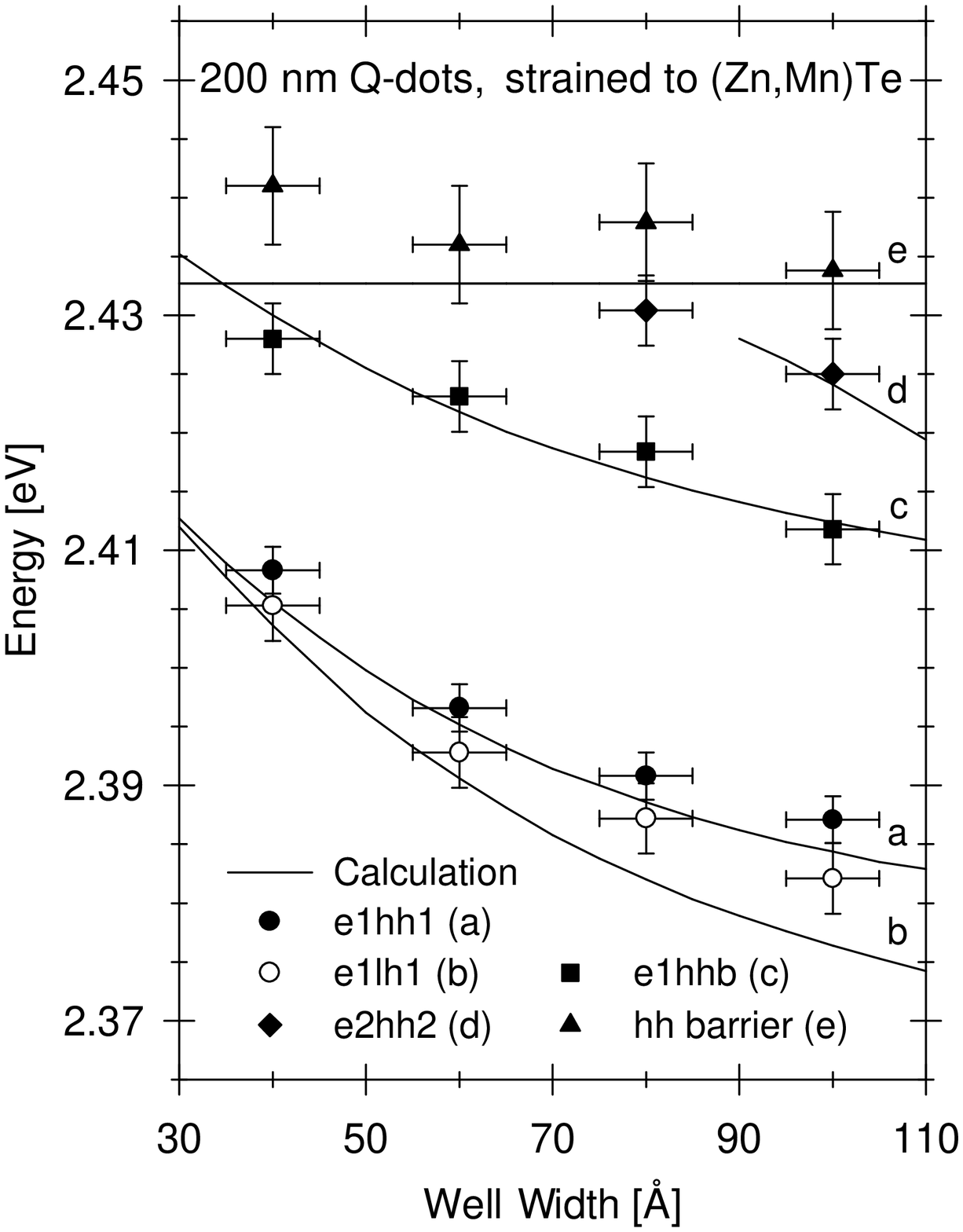,width=0.9\textwidth}
\end{center}
\caption{\label{fig7}Comparison of the calculations (full curves
labelled a to e) for 
the case in which 
the heterostructure is strained to the Zn$_{0.93}$Mn$_{0.07}$Te 
lattice constant with the 
experimental results (symbols) obtained for the excitonic 
transition energies in the four 
200\,nm diameter quantum dot samples each based on a 
Zn$_{0.93}$Mn$_{0.07}$Te/ZnTe multiple 
quantum well sample with a different well width. The horizontal 
error bars result from the 
assumption of an uncertainty of 5\,\AA{} (2 monolayers) in the 
quantum well width and the 
vertical bars represent the fitted linewidth of each transition. }
\end{figure}
The full curves in figure 6 
are calculated assuming a potential profile where the sample is 
strained to the ZnTe lattice 
constant (figure 2, centre). The two horizontal lines indicate the 
hh barrier and lh barrier 
transitions. The other lines follow the calculated quantum well 
width dependences for the 
e2hh2, e1hh1 and e1lh1 transitions and for the e1hhb transition. 
The dependence on 
quantum well width of the e2hh2 transition was only calculated 
for wells wider than 85\,\AA{}, 
for which the second electron state is confined and our transfer 
matrix method is 
applicable. The calculated transition energies of the e1hhb do not 
include a binding energy 
contribution. This contribution should only lead to a reduction of 
the calculated value by a 
few meV because it is a spatially-indirect transition. The error 
bars on the experimental 
values are the fitted linewidths of the transitions (along the 
energy axis) and a 5\,\AA{} error 
bar on the well width axis representing a possible two-monolayer 
deviation from the 
nominal quantum well width. The agreement between the calculated 
values and the fitted 
values for the MQWs is very good especially for the two barrier 
transitions and the two 
lowest quantum well transitions, for which the deviations are 
within the experimental error 
bars. The agreement is not as good for the e2hh2 transition and 
the e1hhb transition, but 
the calculations show the same trend as the data. Figure 7 shows 
the results for the Q-dot 
samples; on this figure, the error bars were determined in the 
same way as for Figure 6. 
We do not observe an increased linewidth for the excitonic 
transitions in the Q-dots, thus 
there is no indication from the BPR data for an inhomogeneous 
strain distribution in the 
Q-dot pillars. Again, we obtain good agreement between theory and 
experiment. There are 
two trends as a function of quantum well width which need to be 
considered. Firstly, the 
experimental values for the barrier transition are slightly higher 
than the calculated values 
for the three wider quantum wells whereas in the case of the MQWs 
we obtained very 
good agreement for the three samples. Secondly, the deviation 
of the experimental results 
from the calculated dependence increases with increasing quantum 
well width for the 
e1hh1 and especially for the e1lh1 transition. The experimental 
data lie above the 
calculated energies and the e1lh1-e1hh1 splitting is less than 
calculated. These trends can 
be understood assuming that the Q-dots are more or less freestanding, 
thus the actual 
lattice constant in the structure approaches an intermediate 
value of the ZnTe lattice 
constant and Zn$_{0.93}$Mn$_{0.07}$Te lattice constant weighted by 
the overall thickness of the 
corresponding layers in the dot pillar:
\begin{equation}
a_\mathrm{par}=\frac{t_{\rm ZnTe}a_{\rm ZnTe}+t_{\rm
   ZnMnTe}a_{\rm ZnMnTe}}{t_{\rm ZnTe}+t_{\rm ZnMnTe}}
   =a_{\rm ZnMnTe}-c\,\Delta a,
\end{equation}
where $t_{\rm ZnTe}$ and $t_{\rm ZnMnTe}$ are the overall 
thicknesses of the 
respective materials, $a_{\rm ZnTe}$ 
and $a_{\rm ZnMnTe}$ are the equilibrium lattice constants of the 
ZnTe well material and the 
Zn$_{0.93}$Mn$_{0.07}$Te barrier material at the measurement
temperature and 
$\Delta a = a_{\rm ZnMnTe} - 
a_{\rm ZnTe}$ and $c = t_{\rm ZnTe}/(t_{\rm ZnTe} + t_{\rm ZnMnTe})$.
Because the depth of etching into the ZnTe 
buffer is uncertain, the exact height of the dots is not known and 
it is difficult to estimate 
the overall thicknesses $t$ in the Q-dot pillars. We therefore 
replace the thicknesses $t$ in 
equation~(8) by the thicknesses of an individual well and barrier, 
ignoring the role of the 
buffer layer. Since it is likely that we have not etched deep into 
the ZnTe buffer layer (as is 
indicated by the additional features corresponding to signals from 
an unetched well in the 
case of the Q-dot sample with 60\,\AA{} well width), this is
justified. We thus obtain 
$c_{40} = 0.2$ 
and $c_{100} = 0.4$; a small value of $c$ indicates that $a_\mathrm{par}$
approaches 
$a_{\rm ZnMnTe}$, so that this 
calculation confirms those Q-dot samples with narrow quantum wells 
are described well 
by the potential situation assumed in the calculation of the
transition energies. For the 
Q-dot samples with wider wells the potential situation will be of 
an intermediate type 
between the two strain states described in figure 2.  Again, it 
should be noted that the 
actual profiles of the electron and the light-hole and heavy-hole 
potentials do not change, 
but are only shifted on an absolute energy scale, varying between 
the two limits given by 
the potential situations depicted in the centre and on the right 
side of figure~2. 

We can summarize the experimental results in the following way: 
the MQW samples are to 
a good approximation strained to the ZnTe lattice constant whereas the
pillars in the 
Q-dot samples are more or less freestanding, which can be understood 
by considering the 
structure of our samples. Similar results have been obtained for 
nanostructures prepared 
from other material systems. For example, a strain relaxation has 
been observed by 
photomodulated reflectivity in quantum wires with wire widths 
between 15\,nm and 
500\,nm etched by RIE on a p$^+$-Si/Si$_{0.8}$Ge$_{0.2}$ 
heterostructure [30] or by x-ray 
diffraction in quantum dots etched from Si$_{1-x}$Ge$_x$ epilayers
with 
$x = 0.038$ and $x = 0.053$ 
on a Si-substrate where the pillars were etched through the epilayer 
into the Si-buffer layer 
[25]. In the III-V materials,  quantum dots of diameter between 230\,nm
and 500\,nm were 
prepared by RIE from a GaAs/Ga$_{0.7}$Al$_{0.3}$As superlattice on a 
GaAs-substrate an 
increasing strain relaxation with decreasing dot diameter was observed [27] by 
photomodulated reflectivity. Finally, in a quantum wire 
sample with wire width of 145\,nm 
prepared from a InAs/GaAs superlattice on a GaAs substrate, 
the superlattice structure in 
the individual wires turned out to be partially relaxed [24].

\section{Conclusions}

We have prepared large-area high-density patterns of 
quantum dots with a diameter of 
200\,nm from a series of four Zn$_{0.93}$Mn$_{0.07}$Te/ZnTe 
multiple quantum well structures of 
different well widths (40\,\AA{}, 60\,\AA{}, 80\,\AA{} and 100\,\AA{}) 
by electron beam lithography followed by 
ion beam etching with Ar$^+$ ions. Examination under a scanning 
electron microscope 
reveals that the patterns are very homogeneous and have a ratio 
of dot area to etched area 
of 1:3. The individual dot pillars have inclined sidewalls with 
an inclination angle of about 
25$^\circ$. We have performed below-bandgap photomodulated 
reflectivity experiments on the 
series of quantum dot samples and, for comparison, on the series of parent 
heterostructures. An analysis of the spectra gives the following
results. 
Firstly, we do not 
see additional confinement effects due to the reduction of the 
dimensionality from two 
dimensions to zero dimensions. Secondly, we observe an increasing 
oscillator strength for 
the light-hole exciton in the quantum dots. A possible explanation 
is that the probe light 
can penetrate into the sidewalls of the quantum dot pillars and 
can thus induce an electric 
dipole moment parallel to the growth direction of the heterostructure. 
This results in an 
increased matrix element for the light-hole exciton whereas the 
matrix element of the 
heavy-hole exciton is not changed. Thirdly, the main changes 
in the excitonic energies as a 
consequence of the nanofabrication process are due to a strain
relaxation 
in the quantum 
dots. We have compared the energy positions of the excitonic 
transitions obtained by 
fitting the lineshapes of the experimental spectra with the results 
of model calculations 
based on a one-dimensional transfer matrix method followed by 
an excitonic binding 
energy calculation. These calculations confirm that the quantum 
dots are freestanding 
whereas the original heterostructures are to a good approximation 
strained to the ZnTe 
buffer.

Our analysis also shows that below-bandgap photomodulated reflectivity, like 
conventional photomodulated reflectivity, is a powerful tool to study 
the effects of 
nanostructure fabrication on semiconductor structures. In this context, 
it should be noted 
that modifications due to the fabrication process can already be 
observed for structural 
dimensions larger than those for which additional confinement effects
occur. 
It is therefore 
of great importance to study nanostructures in this intermediate 
size regime because a 
fundamental understanding of the smaller structures will only be 
attained if a separation of 
the macroscopic effects (strain relaxation, damage) and the 
additional confinement effects 
is possible.

In the future, the effects of the nanostructure fabrication on these 
samples will be 
investigated via magneto-optic experiments which make use of the 
unique magnetic 
properties of dilute magnetic semiconductors. The changes induced 
by the strain 
relaxation in the quantum dots offer interesting possibilities 
in this context. For example, 
the photoluminescence of the quantum dot samples has light-hole 
character in the absence 
of a magnetic field whereas the photoluminescence of the 
heterostructures has heavy-hole 
character. By applying a magnetic field, the character 
of the photoluminescence in the 
quantum dots can be changed from light-hole to 
heavy-hole due to fact that the magnetic 
field splitting of the heavy holes is greater 
than that of the light holes.

\section*{Acknowledgements}

We thank J. J. Davies for a careful reading of the manuscript. 
The support of the 
Engineering and Physical Sciences Research Council of the UK 
under contracts GR/H 
57356 and H 93774 is gratefully acknowledged. 
P. J. Klar thanks the University of East 
Anglia for the provision of a research studentship. 
T. Henning is grateful to the Deutscher 
Akademischer Austauschdienst for a research student grant 
within the framework of the 
HSP II scheme. We are also grateful to the 
Swedish Nanometer Laboratory for the 
possibilty to use their facilities for the preparation of the Q-dot samples.

\end{document}